\documentclass[%
 reprint,
amsmath,amssymb,
aps,
]{revtex4-2}

\usepackage{graphicx}
\usepackage{dcolumn}
\usepackage{bm}
\usepackage[mathlines]{lineno}
\usepackage[colorlinks,linkcolor=blue,anchorcolor=blue,citecolor=red]{hyperref} 
\usepackage{makecell}
\usepackage{mhchem}
\begin{document}


\title{Lattice distortion tuning resistivity invar effect in high entropy alloys}

\author{Hao Chen}
\thanks{These authors contributed equally to this work.}

\author{Yuanji Xu}
\thanks{These authors contributed equally to this work.}
\author{Lihua Liu}

\affiliation{%
Institute for Applied Physics, University of Science and Technology Beijing, Beijing 100083, China 
}%

\author{Yue Chen}
\affiliation{
Department of Mechanical Engineering, The University of HongKong, Pokfulam Road, Hong Kong
}%

\author{Jan Wróbel}%
\affiliation{Faculty of Materials Science and Engineering, Warsaw University of Technology, Woloska 141, 02-507 Warsaw, Poland}

\author{Daoyong Cong}%
\affiliation{State Key Laboratory for Advanced Metals and Materials, University of Science and Technology Beijing, Beijing, China}

\author{Fuyang Tian}
\email{fuyang@ustb.edu.cn}
\affiliation{
Institute for Applied Physics, University of Science and Technology Beijing, Beijing 100083, China 
}%
\affiliation{
Department of Mechanical Engineering, The University of HongKong, Pokfulam 999077, Hong Kong
}%

\author{Yang Wang}
\affiliation{
Pittsburgh Supercomputing Center, Carnegie Mellon University, Pittsburgh, Pennsylvania 15213, USA 
}

\date{\today}

\begin{abstract}
Materials with an ultra-low temperature coefficient of resistivity are desired for the temperature and flow sensors in high-precision electronic measuring systems. In this work, the Kubo-Greenwood formula, implemented in ab initio molecular dynamics simulations, is employed to predict the finite-temperature resistivity of multi-component alloys with severe lattice distortion. We observe a tiny change in resistivity over a wide temperature range in high-entropy alloys. The electronic resistivity invar effect in B2 Ni$_{25}$Co$_{25}$(HfTiZr)$_{50}$ Elinvar alloys results from a balance between intrinsic and residual resistivity. This effect is associated with atomic displacements from ideal lattice sites, which are caused by lattice thermal vibrations and chemical disorder-induced lattice distortions. It is further evidenced by a decrease in lattice distortion with temperature and changes in the electronic density of states.
\end{abstract}

\keywords{Resistivity invar effect; Lattice distortion; Kubo formula; AIMD}
\maketitle

\section{\label{sec:level1}Introduction}
Electrical conductivity or resistivity, is one of the most significant physical properties, of both theoretical and practical importance. As early as 1864, Matthiessen proposed an empirical law for metallic resistivity:  $\rho=\rho_i+\rho_L(T)$, 
where $\rho_L$ is the intrinsic resistivity due to lattice thermal vibrations, and $\rho_i$ is the residual resistivity caused by electron scattering from static defects or impurities. Typically, electronic resistivity increases with temperature above ambient conditions, primarily due to enhanced lattice vibrations. 

Modern technologies, however, demand accurate and stable electrical signals under exetreme conditions which strongly rely on stable resistivity. While materials like Constantan (Cu-Ni) and Manganin (Cu-Mn-Ni), known for their low temperature coefficient of resistance (TCR), have been used for over a century, the development of novel materials with constant resistivity across a wide temperature range remains a significant challenge. To address this, it is essential to evaluate the respective contributions of intrinsic and residual resistivity to the temperature-dependent resistivity of alloys, leveraging current experimental techniques and state-of-the-art first-principles calculations. 

In 1973, Mooij pointed out that the low TCR in disordered alloys with transition metals could be attributed to the very short electron mean free path in these materials \cite{Mooij1973}. Forty years later, Chen $et$ $al.$ \cite{Chen2012} reported a near-constant TCR over a wide temperature range (for instance from 4.2 K to room temperature) in certain high-entropy alloys (HEAs) \cite{Cantor2004,Yeh2004}, highlighting their unique properties within this fascinating field. The factors influencing the electronic mechanisms in HEAs were discussed, including spin dependence, chemical disorder, high density of lattice defects, weak phonon effects, $s$-$p/d$ scattering in transition metals, phonon drag effects at low temperature, and the Mott-Ioffe-Regel (MIR) limit \cite{Chen2012,Shafeie2015,Mu2019,Mu2019a,Raghuraman2021,Uporov2022,Bag2021,Kota2009,Fradin1974}. These effects are modeled phenomenologically, for instance, by the equation $\rho(T) = a_0 + a_1 T + a_2 T^2$ at high temperatures, where $a_0$, $a_1$, and $a_2$ are fitting parameters related to electron-phonon or electron-electron scattering \cite{Dong2023}. For HEAs with B2 phase, the equation $\rho(T) = \rho_0 + A(T) \ln(T) + B(T) T^2 + C(T) T^3 + D(T) T$ was used, where $\rho_0$ remains nearly constant across different temperature ranges, while the coefficients $A(T)$, $B(T)$, $C(T)$, and $D(T)$ decrease with increasing temperature \cite{Chen2012}. However, quantitative reasoning based on intrinsic and residual resistivity is still lacking.

The strong scattering of electrons by lattice distortion in HEAs leads to high resistivity \cite{Yeh2016}. Mu $et$ $al.$ found that the scattering mechanism due to ideal chemical disorder is relatively weak in fcc HEAs \cite{Mu2019}, while it is strong in some bcc HEAs \cite{Mu2019a}. The electrical resistivity of refractory HEAs is generally high \cite{Dong2023}, due to the severe lattice distortion in bcc structures caused by chemical disorder \cite{Song2017}. Given the significant chemical disorder and the associated lattice distortion \cite{Bonfanti2024}, HEAs are expected to serve as excellent model systems for studying defect-induced residual resistivity at finite temperature and for further tuning of the low TCR. Note that, in the present work, defects in HEAs are associated with chemical disorder (compared to ideal order) and its induced lattice disorder (relative to perfect lattice).
To computationally study electronic scattering in alloys, the semi-classical Boltzmann transport equation (BTE), Kubo’s linear response formula, and the nonequilibrium Green’s function methods have been used within the framework of first principles \cite{Mu2018}. The Kubo-Greenwood (KG) formula, in combination with multiple scattering theory, specifically the Korringa-Kohn-Rostoker method coupled with the coherent potential approximation (KKR-CPA), has been used to calculate the residual resistivity of HEAs \cite{Raghuraman2021,Raghuraman2023,Mu2019,Mu2019a,Dong2023}. Using the linearized Boltzmann equation implemented in BoltzTraP2 \cite{Madsen2018}, Wang $et$ $al.$ discussed the electrical transport properties of Si-Ge-Sn medium entropy alloys with semiconductor performance \cite{Wang2020a}. However, to our knowledge, little work has been done on how temperature-dependent lattice distortion and thermal vibrations influence electronic resistivity in HEAs with significant chemical disorder and severe lattice distortion.

In this work, we find that the resistivity invar effect (extremely low TCR) in single-crystalline 25Co-25Ni-16.67Hf-16.67Ti-16.67Zr alloys, as prepared experimentally \cite{He2022}, results from the decrease in lattice distortion with temperature. For simplicity, this HEA is denoted as \ce{Ni25Co25(TiHfZr)50} (in atomic percentage). We first apply the KG formula combined with $ab$ $initio$ molecular dynamics (AIMD) simulations to accurately calculate the temperature-dependent resistivity of solid alloys, although this method has been applied to melting metals and warm dense matter \cite{Pozzo2011,Knyazev2013,Vlvcek2012,Svensson2023,Calderin2017,Liu2021}. More importantly, we quantify lattice distortion, thermal vibrations, and the electronic density of states (DOS) to illuminate the physical mechanism behind the resistivity invar effect in HEAs.

\section{Methodology}
Theoretical calculations were performed using density functional theory (DFT) \cite{Hohenberg1964,Kohn1965} and the projector augmented wave (PAW) method, as implemented in the Vienna $ab$ $initio$ simulation package (VASP) \cite{Kresse1996CMS,Kresse1996PRB,Blöchl1994}. The Perdew, Burke, and Ernzerhof (PBE) exchange-correlation functional \cite{Perdew1996} was used, with a plane-wave energy cutoff of 400 eV for all calculations. 

We computed the resistivity in the static limit for each ionic configuration using the KG formula \cite{Kubo1957} implemented in VASP \cite{Chen2024}. The KG formula is given as
\begin{align}\label{eq1}
\sigma(\omega) &=\frac{2\pi e^2\hbar^2}{3m^2_e{V}\omega}\sum_\mathbf{k}\sum_{i,j=1}^N\sum_{\alpha=1}^3 W(\mathbf{k})[F(\epsilon_{i,\mathbf{k}})-F(\epsilon_{j,\mathbf{k}})] \notag \\ 
&\quad \times |\langle \Psi_{j,\mathbf{k}} | \nabla | \Psi_{i,\mathbf{k}} \rangle|^2 
 \delta(\epsilon_{j,\mathbf{k}}-\epsilon_{i,\mathbf{k}} - \hbar \omega),
\end{align}
where $\sigma$ represents electrical conductivity and $\rho$ is the resistivity, given by $\rho = 1/\sigma$. $e$ and $m_e$ represent the electron charge and mass, $V$ is the volume, and $\omega$ is the frequency. $W(\mathbf{k})$ represents the weight of the $\mathbf{k}$-point, and $F$ denotes the Fermi-Dirac distribution function. In practical simulations, the finite volume of the system leads to discrete eigenvalues, necessitating the broadening of the $\delta$ function. To address this, we employ a Gaussian broadening scheme. The width of the Gaussian is carefully chosen to be sufficiently small to avoid recovering local oscillations in the conductivity, which would otherwise arise due to the discrete nature of the band structure \cite{Desjarlais2002}.

The finite-temperature equilibrium configurations are extracted from AIMD simulations at equal time intervals, as described below. We then calculate the electronic band structures of all configurations, obtaining the energy eigenvalues and the squared modulus of the transition matrix for each configuration. We further average the Onsager coefficients of all configurations, obtaining the temperature dependence of the resistivity. We note that the KG formula has also been implemented in VASP by Calder and Knyazev $et$ $al.$ \cite{Calderin2017,Knyazev2013}, but to the best of our knowledge, these schemes have not been extended to alloys.

To perform AIMD simulations and describe the chemical disorder of HEAs, we used the similar atomic environment (SAE) method to generate random solid-solution configurations \cite{Song2017,Tian2020}. The number of atoms in SAE supercells for the simple-phase bcc TiZrHfNb, bcc TiZrHfNbTa, and B2 \ce{Ni25Co25(TiHfZr)50} HEAs, as found in experiments, are 128, 160, and 96, respectively \cite{Braic2014,Senkov2011,He2022}. We constructed the configurations for \ce{Ni25Co25(TiHfZr)50} with partial order, where some Zr atoms at corner sites exchange with Co and Ni atoms at center sites, as described in experiments \cite{He2022}. To construct the objective function in the SAE method, we choose cutoff radii of 7-times and 1.6-times the nearest-neighbor interatomic distance for two-site and three-site clusters, respectively. Although different objective functions are adopted, the SAE method is comparable with the special quasi-random structure (SQS) method \cite{Tian2020}.

Initially, we performed an NVT ensemble (constant number of atoms, constant volume, constant temperature), using a Nose-Hoover thermostat for temperature control. Each thermodynamic state was simulated for 5000 time steps, with 2 fs per step. After equilibration, we took snapshots of nuclear positions every 200 MD steps. 
For each snapshot, we performed static electronic DFT calculations and used the results to evaluate the Onsager coefficients derived from the KG formula. For smearing, we used the Fermi-Dirac function with an electronic temperature corresponding to the system temperature. The approximated expression of the  $\delta$ function is crucial in practice for evaluating the Onsager coefficients \cite{Bulanchuk2021,Subedi2022}. 

We adopted the Debye model \cite{Lonsdale1948,Zhou2024} to estimate the atomic displacement induced by thermal vibrations, relative to ideal lattice sites. We used the temperature-dependent effective potential (TDEP) method \cite{Hellman2011} to extract the finite-temperature phonon density of states from AIMD simulations and obtained the necessary input parameter $\Theta_D$ (Debye temperature). For comparison, the KG formula, implemented within the Multiple Scattering Theory (MuST) package \cite{MuST}, is also employed to calculate the residual resistivity due to ideal chemical disorder. MuST is an $ab$ $initio$ electronic structure calculation software suite, with petascale computing capability for disordered materials. Here, the coupled coherent potential approximation (CPA) accurately describes the ideal chemical disorder. Readers are referred to references~\cite{Raghuraman2021,Raghuraman2023} for the calculation details.

\section{Results and discussion}
\begin{figure}[htbp]
\includegraphics[width=\linewidth]{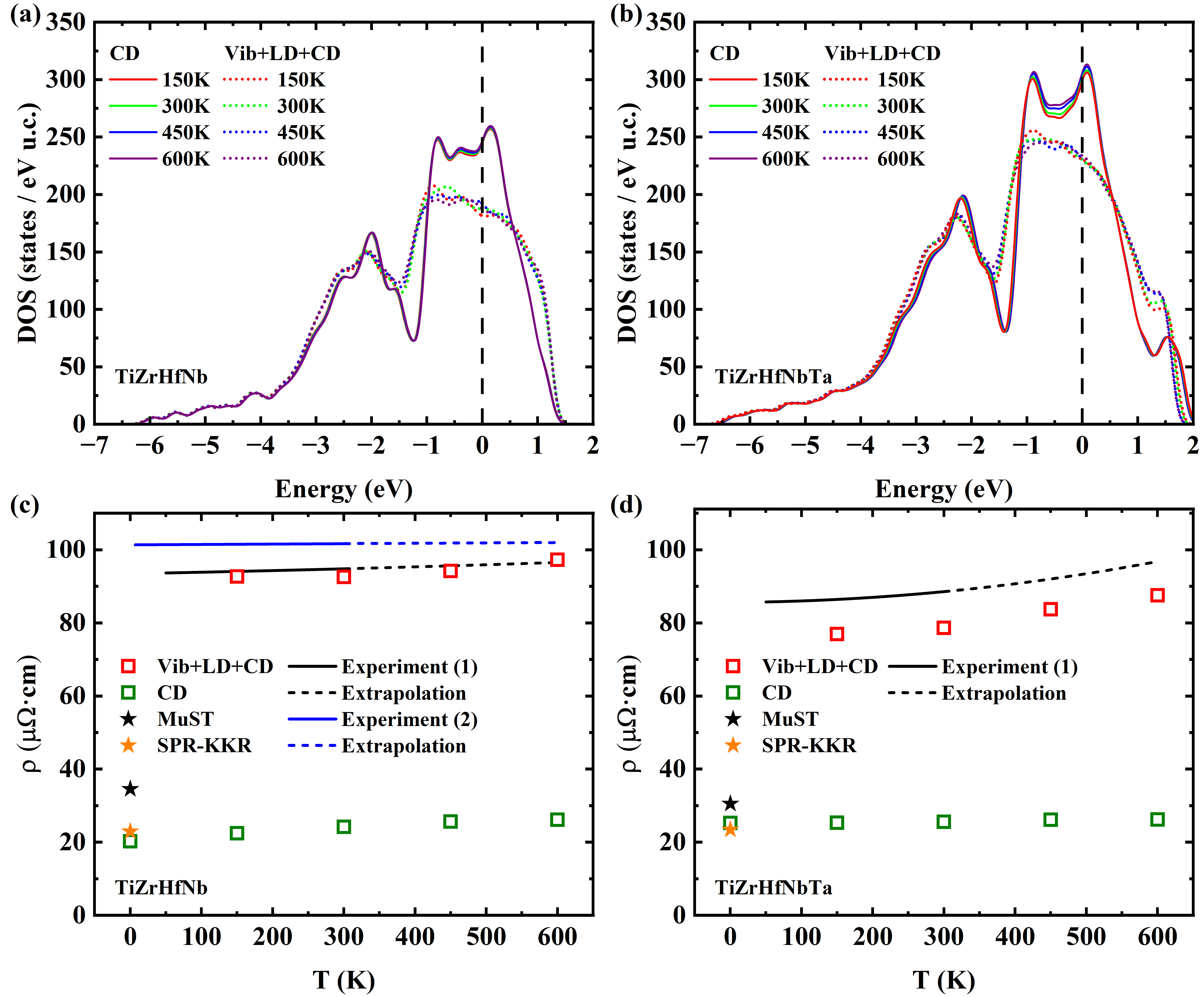}
\caption{\label{RHEA} Electronic density of states (DOS) and resistivity of TiZrNbHf (a,c) and TiZrNbHfTa (b,d) versus temperature from our KG-AIMD and MuST calculations, the published SPR-KKR calculations~\cite{Dong2023}, as well as experiments~\cite{Dong2023,Uporov2022}. CD (green square) is the resistivity from the ideal chemical disorder, LD the resistivity from CD induced lattice distortion, and Vib the resistivity from thermal vibration. Total resistivity (red square) includes the contributions from thermal vibration and defects (Vib+LD+CD).}
\end{figure}

In classical models, electrical resistance can be explained by treating electrons and phonons (crystal lattice) as wave-like entities. When the electron wave propagates through the lattice, wave interference leads to resistance. The more regular the lattice, the fewer disturbances, resulting in lower electrical resistivity. Recall the Drude model: $\frac1{\rho} = \frac{ne^2\tau}{m^*}$, where $\tau$ denotes the relaxation time (the average time between collisions), $n$ is the charge carrier density, and $m^*$ is the effective electron mass. The charge carrier density $n$ can be estimated from the Fermi-Dirac distribution of the electronic density of states (DOS). Figures \ref{RHEA} (a) and (b) show the electronic DOS for ideal chemical disorder (CD) and the AIMD simulations, which include CD, lattice thermal vibration (Vib), and lattice distortion (LD, describing atomic displacement from ideal lattice sites in HEAs). The electronic DOS at different temperatures are similar for TiZrHfNb and TiZrHfNbTa. As Hunklinger pointed out, the charge carrier density is nearly independent of temperature due to the Fermi distribution \cite{Hunklinger2014}. Thus, the only temperature dependence arises from $\tau$. A regular lattice leads to a long $\tau$ (mean free path of electrons), resulting in low resistivity for perfect metals. Two primary scattering sources affecting $\tau$ and thus influencing resistance are lattice thermal vibration and defects. 

Thermal lattice vibrations, which cause atomic positions to deviate from the ideal positions associated with the underlying crystal lattice, grow stronger with increasing temperature and mostly increase the volume of material. Due to the different atomic radii of the alloying elements, larger atoms push away neighboring atoms, while smaller ones have extra space around them. No atom resides perfectly on the ideal lattice site, and atomic size mismatch gives rise to considerable lattice distortions. Additionally, interatomic charge transfer can lead to short-range order and lattice distortions, even when atomic radii are similar. This implies that chemical disorder induces the loss of long-range order in crystalline materials, strongly affecting the coherence of electronic waves diffracted by adjacent ions.

Unlike the KG formula coupled with CPA, the combination of the KG formula and supercell is limited by size effects. At low temperatures, the electron mean free path of simple metals is relatively long, requiring large supercells for accurate resistivity simulations. For example, Subedi $et$ $al.$ constructed a 256-atom fcc supercell of aluminum to calculate its resistivity over a temperature range of 50 K to 450 K. However, the results with a smearing width of 0.02 eV were still about 2.4 times larger than experimental values \cite{Subedi2022}. When using a smaller smearing width of 0.005 eV, the simulated resistivity is closer to experimental values.

However, the KG formula has been effectively applied to melting metals and warm dense matter due to their high disorder and the resulting short mean free path of electrons. Simulated resistivity often agrees well with experimental results, even when small supercells are used \cite{Liu2021,Migdal2019}. In HEAs with significant chemical disorder, strong electron scattering from disorder-induced lattice distortion, rather than lattice thermal vibrations, dramatically reduces the electron mean free path to nearly the interatomic spacing. Therefore, it is reasonable to apply the KG-AIMD combination, even using small-size supercells, to investigate the resistivity of HEAs and its dependence on temperature.

To investigate the effect of ideal chemical disorder (CD) on resistivity, we adopted the ideal lattice structure (with alloying elements placed on ideal lattice sites) to calculate the resistivity at finite temperatures based on the equilibrium volume from AIMD simulations. First, we calculate the residual resistivity at 0 K. Our resistivity calculations from ideal CD (green squares in Figs. \ref{RHEA}(c,d)) are consistent with the CPA results (stars in Fig. \ref{RHEA}(c,d)) \cite{Dong2023}, except for the MuST results for TiZrHfNb. This suggests that the SAE method can accurately describe ideal CD using a small supercell, and the KG formula can effectively capture the residual resistivity due to ideal CD. From Fig. \ref{RHEA} (c, d), we observe that the resistivity from CD is very small, contributing only 20-25\% to the total resistivity, which includes contributions from CD, lattice vibrations (Vib), and CD-induced lattice distortion (LD). While lattice thermal expansion is considered at finite temperatures, the resistivity from ideal CD shows a slight increase for TiZrHfNb and remains nearly constant for TiZrHfNbTa.

We note that the peak of the electronic DOS at the Fermi level is shown in Fig. \ref{RHEA} (a, b) for the CD case. When including the contributions from thermal vibration and CD-induced lattice distortion, there is a significant reduction in DOS at the Fermi level. This corresponds well to the phenomenon where the combination of thermal vibration and lattice distortion leads to a substantial increase in total resistivity, as shown in Fig. \ref{RHEA} (c, d). Since the electronic DOS at the Fermi level for TiZrNbHf and TiZrNbHfTa is relatively sharper in the CD cases, the thermal vibration and CD-induced lattice distortion effectively smear the Fermi energy, lowering the DOS near the Fermi level, and thus increasing the resistivity. These results support the feasibility of our KG-AIMD approach for studying electronic transport in HEAs.

Furthermore, we compared our KG-AIMD simulations on the total resistivity of HEAs with available experimental measurements. In Fig. \ref{RHEA}(c, d), the experimental temperature ranges are 55-300 K (Dong $et$ $al.$ \cite{Dong2023}) and 6.4-300 K (Uporov $et$ $al.$ \cite{Uporov2022}) for TiZrHfNb, and 50-300 K for TiZrHfNbTa \cite{Dong2023}. Based on their fitting parameters \cite{Dong2023,Uporov2022}, we extrapolated the temperature to 600 K (dashed lines in Fig. \ref{RHEA} (c, d)). Despite slight differences between the two experimental datasets, both show minimal temperature dependence. Our KG-AIMD simulations (red squares in Fig. \ref{RHEA}) are consistent with these experimental results. The temperature coefficient of resistivity is given by TCR$= \frac{1}{\rho}\frac{d\rho}{dt}$. For TiZrHfNbTa, our simulated resistivity is slightly lower than the experimental values, but the trend of resistivity change with temperature agrees well with the experiments. Our simulated TCR at 600 K is $0.49 \times 10^{-3}$ K$^{-1}$, which is close to the experimental value of $0.36 \times 10^{-3}$ K$^{-1}$.
\begin{table}[htbp]
\caption{\label{tab:table1}%
Temperature dependence of resistivity with fitting function $\rho(\Delta T)=\rho_0(1+\alpha_1 \Delta T + \alpha_2 \Delta T^2)$. $\rho_0$ ($\mathrm{\mu\Omega{\cdot}cm}$) being the resistivity at the reference temperature of 300 K, $\alpha_1$ and $\alpha_2$ being temperature coefficient of resistivity ($1\times10^{-4}/K$, $1\times10^{-7}/K^2$). The experimental values are also listed for typical  alloys. }

\begin{ruledtabular}
\begin{tabular}{lccc}
\textrm{Alloy}&
\textrm{$\rho_0$}&
\textrm{$\alpha_1$}&
\textrm{$\alpha_2$} \cr
\colrule
Ni$_{25}$Co$_{50}$(TiZrHf)$_{50}$(B2) & 129.32&0.49& 0.46\\
Ni$_{25}$Co$_{50}$(TiZrHf)$_{50}$(PO) & 155.99 &-0.74&1.19 \\
TiZrNbHf & 92.62&0.53&3.84 \\
Exp.\cite{Dong2023}&94.83&0.40&0.23\\
TiZrNbHfTa & 79.30 &2.65&2.98 \\
Exp.\cite{Dong2023} &88.59&0.11&3.32\\
Ni$_{60}$Cu$_{40}$&52.57&-7.27&10.44\\
Exp.\cite{Paras2024}&53.84&-0.82&0.61 \\

\end{tabular}
\end{ruledtabular}
\end{table}

To study the trend of resistivity, we used the phenomenologically quadratic polynomial $\rho(\Delta T) = \rho_0(1 + \alpha_1 \Delta T + \alpha_2 \Delta T^2)$, which fits the total resistivity at different temperatures. Table \ref{tab:table1} lists the fitting parameters at room temperature. The reference resistivity $\rho_0$ is in good agreement between calculations and experiments at 300 K. According to Matthiessen’s rule, the resistivity is influenced by electron-phonon ($\sim T$) scattering, quantified by $\alpha_1$, and electron-electron ($\sim T^2$) scattering, quantified by $\alpha_2$. While this approximation is crude, it provides insight into the contributions of different scattering mechanisms. In particular, for classic Ni-Cu alloys, which show a slight decrease in resistivity, electron-phonon scattering contributes negatively, while electron-electron scattering contributes positively.


Experiments report that the single-crystalline Ni$_{25}$Co$_{25}$(HfTiZr)$_{50}$ HEA is a high-performance ultraelastic metal with superb strength, sizeable elastic strain limit, and temperature-insensitive Young's modulus \cite{He2022}. This Elinvar effect is discussed via the decrease of local shear strain and lattice distortion with temperature. Ni$_{25}$Co$_{25}$(HfTiZr)$_{50}$ adopts the B2 structure, where Ni and Co atoms occupy the corner sites, while Ti, Zr, and Hf randomly sit on the body center position. The scanning transmission electron microscope (STEM) observations suggest \cite{He2022} that this alloy constructed with 25\% of the Zr atoms from sublattice A exchanging with Co and Ni atoms on sublattice B matches their experimental samples more closely. Obviously, this configuration is a more severe chemical disorder than B2 Ni$_{25}$Co$_{25}$(HfTiZr)$_{50}$, we name the partial-order configuration as PO. 

\begin{figure}[htbp]
\includegraphics[width=\linewidth]{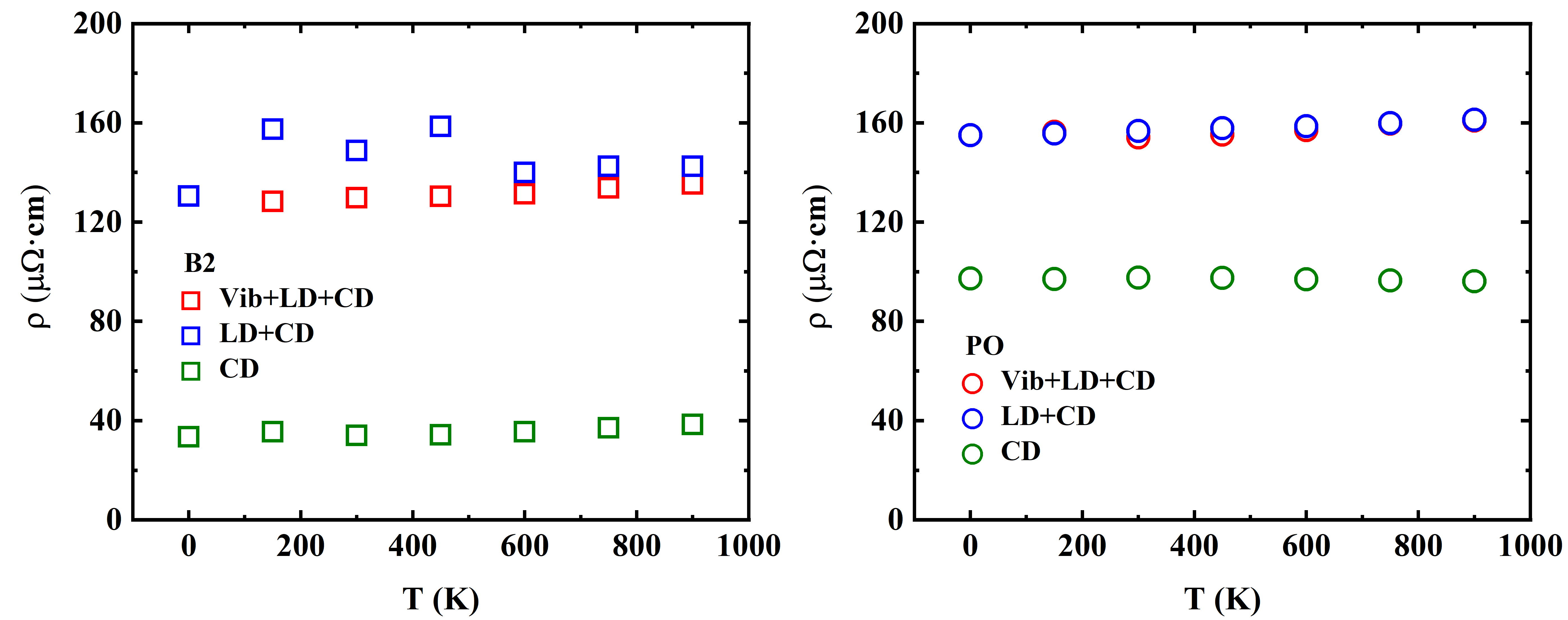}
\caption{\label{invarRHEA} Resistivity of B2 and PO Ni$_{25}$Co$_{25}$(TiHfZr)$_{50}$ with temperature. The ideal chemical disorder (CD), the chemical disorder induced lattice distortion (LD), and thermal vibration coupled with lattice distortion (Vib+LD) are considered as the contributions to resistivity.}
\end{figure}

From Fig.\ref{invarRHEA}, it is evident that severe chemical disorder significantly increases the resistivity of Ni$_{25}$Co$_{25}$(HfTiZr)$_{50}$ HEA, from 35 $\mathrm{\mu\Omega{\cdot}cm}$ (for the B2 structure) to 100 $\mathrm{\mu\Omega{\cdot}cm}$ (for the PO configuration). This suggests that disorder between different sublattices has a significant effect on the residual resistivity. When chemical disorder-induced lattice distortions are taken into account, the resistivity increases further. For the B2 structure, the resistivity rises from 35 to 130 $\mathrm{\mu\Omega{\cdot}cm}$, while for the PO configuration, the resistivity increases from 100 to 160 $\mathrm{\mu\Omega{\cdot}cm}$. Therefore, lattice distortions induced by chemical disorder play a crucial role in the residual resistivity of Ni$_{25}$Co$_{25}$(HfTiZr)$_{50}$ HEA.

Interestingly, when we further consider lattice thermal vibrations and their associated intrinsic resistivity, we observe that the total resistivity of (NiCo)$_{50}$(TiHfZr)$_{50}$ does not continue to increase. Instead, as shown in Fig. \ref{invarRHEA}, the total resistivity either remains constant or is even smaller than the resistivity induced solely by lattice disorder. The intrinsic resistivity increases with temperature, while the degree of ideal chemical disorder remains nearly unchanged across the temperature range. The residual resistivity due to chemical disorder exhibits only a slight increase, as shown in Fig. \ref{invarRHEA} for both B2 and PO configurations of Ni$_{25}$Co$_{25}$(TiHfZr)$_{50}$. Therefore, the lattice distortion-induced residual resistivity appears to be primarily responsible for the observed behavior. Additionally, reduced electron scattering due to lattice distortion with increasing temperature helps to lower the residual resistivity, contributing to the near-constant resistivity change observed in the material.

\begin{figure}[htbp]
\includegraphics[width=\linewidth]{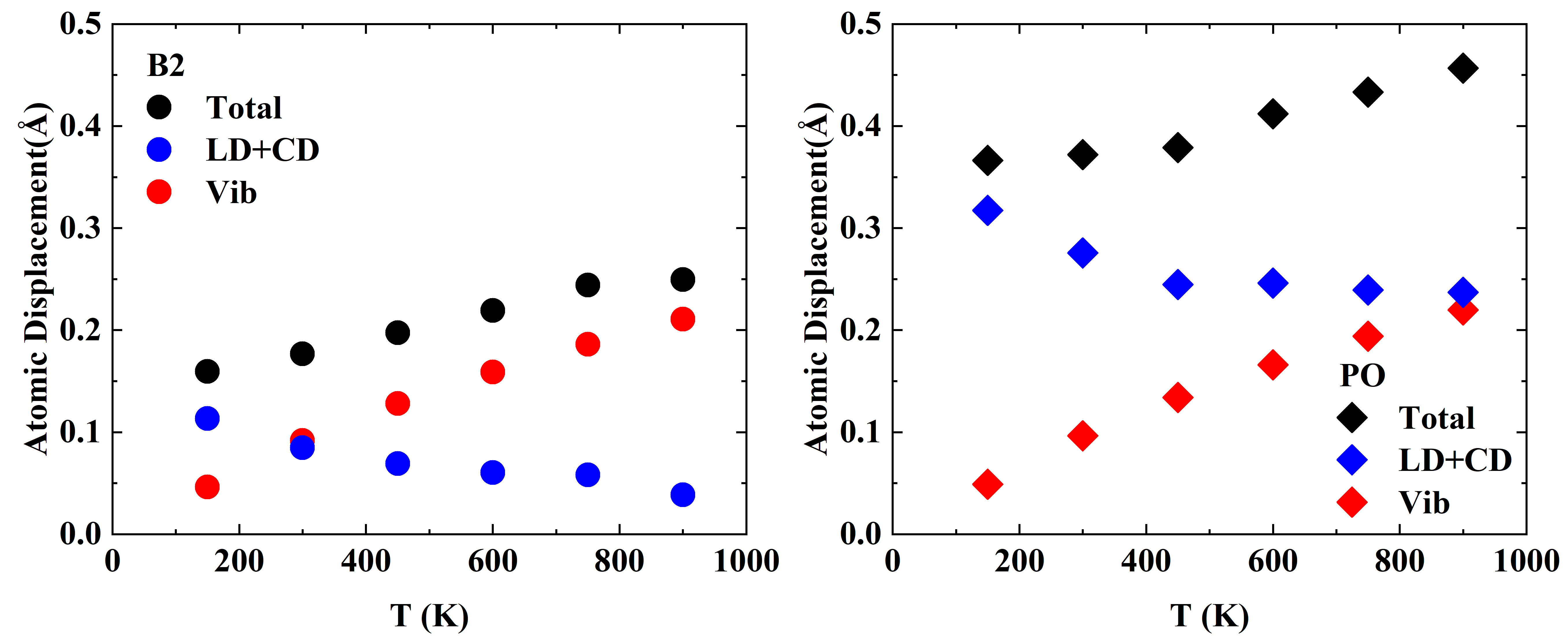}
\caption{\label{fig3} Total and partial atomic displacements away ideal lattice sites from thermal vibration (Vib) and lattice distortion (LD) for B2 and PO \ce{Ni25Co25(TiHfZr)50} with temperature. Total lattice distortion (Total) contains contributions from the thermal vibration and chemical disorder induced lattice distortion.}
\end{figure}

To investigate the change in chemical disorder (CD)-induced intrinsic lattice distortion with temperature, we examine the atomic displacement from ideal lattice sites caused by both thermal vibrations and lattice distortions, as atomic displacement is directly linked to electron scattering. As shown in Fig. \ref{fig3}, although the total atomic displacement due to thermal vibration and lattice distortion increases with temperature, the lattice distortion-induced displacement decreases. This happens because the atomic displacement caused by thermal vibration increases rapidly with temperature, overshadowing the effect of lattice distortion. The reduced lattice distortion leads to lower residual resistivity. Consequently, the opposing trends between the residual resistivity caused by lattice distortion and the intrinsic resistivity due to thermal vibration contribute to the observed resistivity invar effect over a broad temperature range.

\begin{figure}[htbp]
\includegraphics[width=\linewidth]{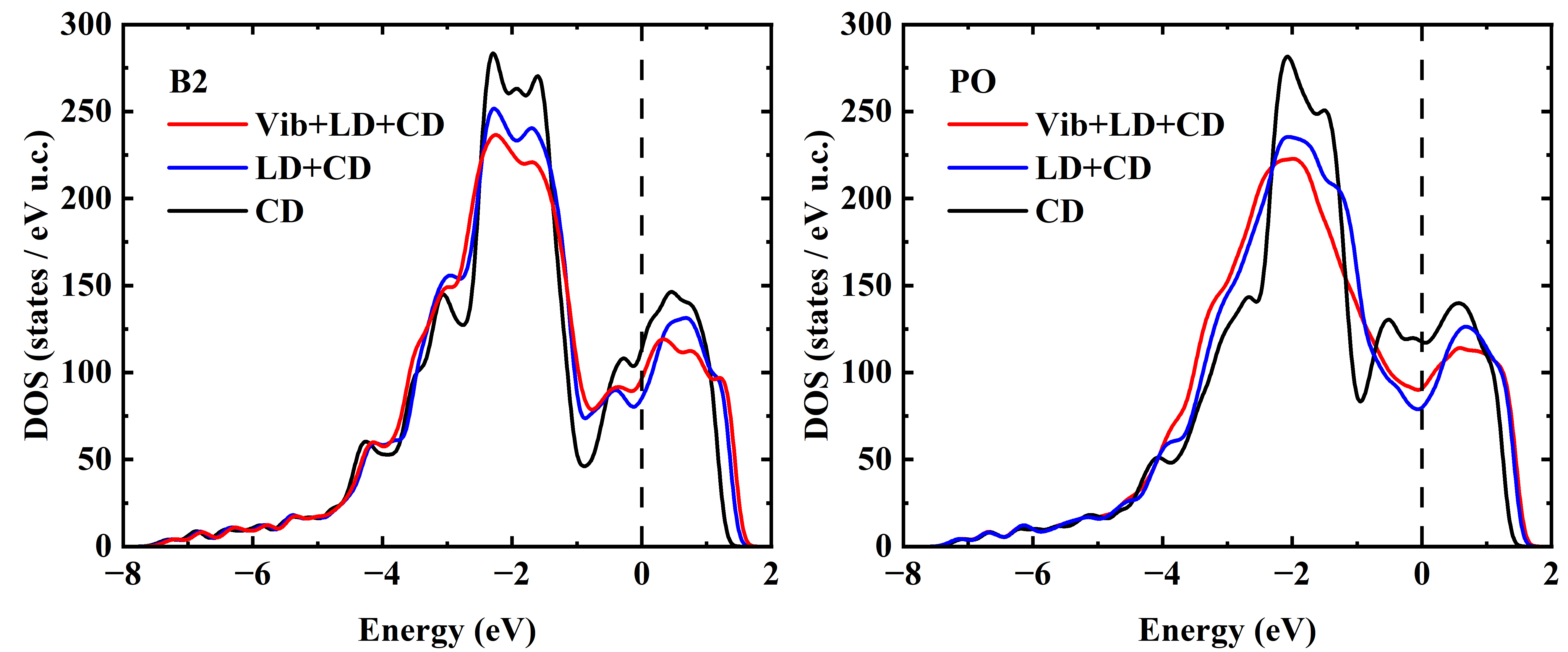}
\caption{\label{fig4} Electronic density of states (DOS) from ideal chemical disorder (CD), lattice distortion (LD) coupled with thermal vibration (Vib) for B2 and PO \ce{Ni25Co25(TiHfZr)50} HEAs at 900 K.}
\end{figure}

The resistivity invar effect is further shown by the electronic DOS shown in Fig. \ref{fig4}. Considering the influence of volume expansion from thermal vibration at 900 K, we calculate the electronic DOS of configurations with ideal chemical disorder and lattice distortion. Comparisons suggest that the DOS near the Fermi level from AIMD configurations (Vib+LD+CD) is slightly larger than that from chemical disorder induced lattice distortion (LD+CD), but smaller than the cases with only chemical disorder (CD). It implies that lattice distortion significantly decreases the DOS at the Fermi level. Because the thermal vibration itself does not make a noticeable change in DOS (see Fig.\ref{RHEA}), for B2 and PO \ce{Ni25Co25(TiHfZr)50}, the thermal vibration decreases the degree of severe lattice distortion from chemical disorder and further increases the DOS at Fermi level, compared to the effect of chemical disorder and its induced lattice distortion. Accordingly, the thermal vibration increases the intrinsic resistivity but indirectly decreases residual resistivity with temperature. 


\section{Conclusion}
In summary, we extended a computational approach, based on a combination of the Kubo-Greenwood formula and $ab$ $initio$ molecular dynamics simulation, to the study of the resistivity of high entropy alloys with short electron mean free paths induced by lattice distortion. Using this scheme, we are able to distinguish the contributions from the chemical disorder, lattice distortions, and lattice thermal vibration to the resistivity via the classical mechanism composed of intrinsic and residual resistivity. Especially the balance of atomic displacements from lattice thermal vibrations and lattice distortion enables the resistivity invar effect to appear in Ni$_{25}$Co$_{25}$(TiZrHf)$_{50}$ HEAs. The degree of chemical disorder is expected to be a powerful way for tuning the resistivity of alloys. The analysis of the degree of lattice distortion and electronic density of states could accelerate the engineering application of B2 Ni$_{25}$Co$_{25}$(TiZrHf)$_{50}$ with both Young's modulus and resistivity invar effects.
\begin{acknowledgments}
We are grateful to Prof. Yandong Wang for giving constructive suggestions. We thank Dr. Xianteng Zhou, Zhen Yang, and Jitong Song for discussion on the temperature-dependent effective potential, thermal vibration model, and experimental results. This work was supported by the National Natural Science Foundation of China (Grant Nos. 52371174 and 12204033).
\end{acknowledgments}


\nocite{*}


%

\end{document}